\documentclass[twocolumn, a4paper, 10pt, showpacs, prl, reprint, aps, superscriptaddress, amsmath, amssymb, amsfonts, floatfix]{revtex4-1}
\usepackage{graphicx, array}
\usepackage{rotating}
\usepackage[caption=false]{subfig}

\begin{document}

\title{Effect of structural relaxation on the electronic structure of graphene on hexagonal boron nitride}
\author{G.J. Slotman}
\affiliation{Institute for Molecules and Materials, Radboud University, Heyendaalseweg 135, 6525AJ Nijmegen, The Netherlands}
\author{M.M. van Wijk}
\affiliation{Institute for Molecules and Materials, Radboud University, Heyendaalseweg 135, 6525AJ Nijmegen, The Netherlands}
\author{Pei-Liang Zhao}
\affiliation{Department of Applied Physics, Zernike Institute for Advanced Materials,University of Groningen, Nijenborgh 4, NL-9747AG Groningen, The Netherlands}
\author{A. Fasolino}
\affiliation{Institute for Molecules and Materials, Radboud University, Heyendaalseweg 135, 6525AJ Nijmegen, The Netherlands}
\author{M.I. Katsnelson}
\affiliation{Institute for Molecules and Materials, Radboud University, Heyendaalseweg 135, 6525AJ Nijmegen, The Netherlands}
\author{Shengjun Yuan}
\email{s.yuan@science.ru.nl}
\affiliation{Institute for Molecules and Materials, Radboud University, Heyendaalseweg 135, 6525AJ Nijmegen, The Netherlands}

\date{\today}

\begin{abstract}
We performed calculations of electronic, optical and transport properties of graphene on hBN with realistic moir\'e patterns. The latter are produced by structural relaxation using a fully atomistic model. This relaxation turns out to be crucially important for electronic properties. We describe experimentally observed features such as additional Dirac points and the "Hofstadter butterfly" structure of energy levels in a magnetic field. We find that the electronic structure is sensitive to many-body renormalization of the local energy gap.
\end{abstract}


\maketitle

%introduction
The physical properties of van der Waals heterostructures can change drastically in comparison with the ones of the constituent two-dimensional materials~\cite{geim2013van}. Recent experiments of graphene on hexagonal boron-nitride (hBN) show that hBN can act like an effective periodic potential for graphene, leading to secondary Dirac points~\cite{xue2011scanning, yankowitz2012emergence}. The graphene/hBN heterostructures are of fundamental interest as an example of a quantum mechanical system with tunable incommensurate potentials. Such incommensurate potentials are important for quasicrystals~\cite{divincenzo1991state} but they are not tunable, whereas in the graphene-hBN systems it is possible to change the potential by changing the mutual orientation of graphene and hBN layers. 
It was long predicted that a system under influence of both a crystal potential and a magnetic field, with a magnetic period incommensurate with that of the crystal, would exhibit a recursive spectrum now called Hofstadter's butterfly~\cite{hofstadter1976energy}, which has been observed in experiments with misaligned graphene on hBN in 2013~\cite{hunt2013massive, ponomarenko2013cloning, dean2013hofstadter}.

Although hBN has a structure similar to that of graphene, the lattice mismatch of $1.8$\% will cause moir\'e patterns, meaning that there is no uniform stacking in the sample. An extra difficulty is posed by the recently observed transition at very small angles from an incommensurate state, with little deformation of graphene, to a commensurate state, where regions of stretched graphene are separated by domain walls~\cite{woods2014commensurate}. The computational challenge lies in the fact that at such angles these superlattices have unit cells consisting of tens of thousands atoms, making it impossible to study them using methods such as DFT. The tight-binding propagation method (TBPM) described in~\cite{yuan2010modeling}, can be used to study systems with hundreds of millions of atoms, circumventing this problem.
In this research, we apply TBPM to graphene-hBN heterostructures with various rotation angles, based on structures determined by atomistic simulations of realistic moir\'e patterns. 

Various approaches have been developed to describe graphene-hBN with an effective tight-binding (TB) Hamiltonian~\cite{wallbank2013generic, diez2014emergence}. While these methods give reasonable results, they lack the flexibility needed to apply them to other systems. We present an approach consisting of three parts, namely: i) structural relaxation of graphene on top of hBN with an empirical potential, ii) modification of the TB parameters due to this relaxation, and iii) calculation of electronic properties with these modified TB-parameters.
There are multiple advantages to this approach. First, the construction of the TB model is solely based on the three-dimensional coordinates of the carbon atoms, thus one could use this same method for graphene on top of other substrates or graphene under mechanical strain. Second, it is easy to incorporate extra disorder such as carbon vacancies, ad-atoms and ripples, etc.
%%image placed here to get it on the second page
\begin{figure*}[t!]
    \includegraphics[width=0.23\linewidth]{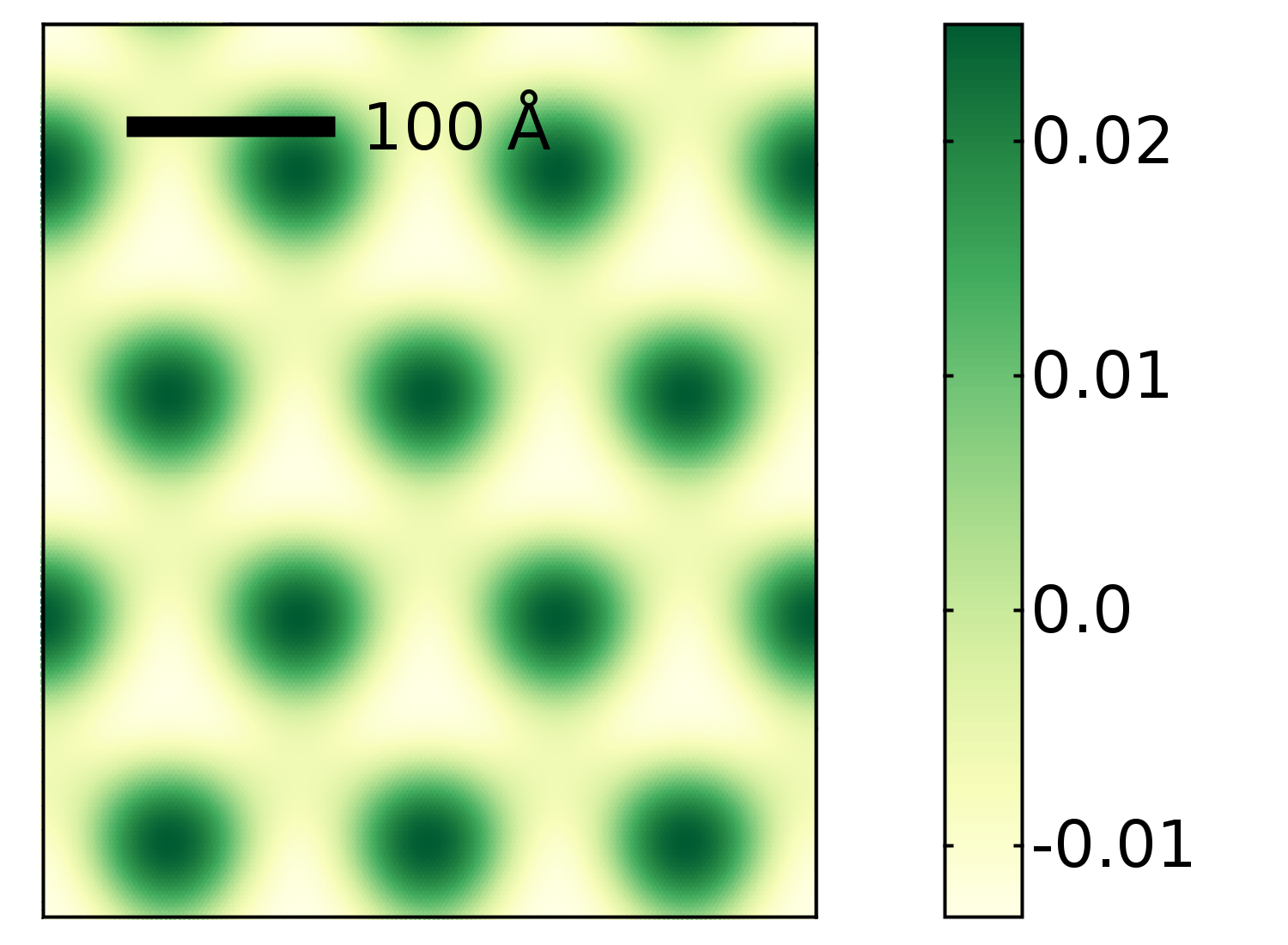}
    \includegraphics[width=0.1533333333333333\linewidth]{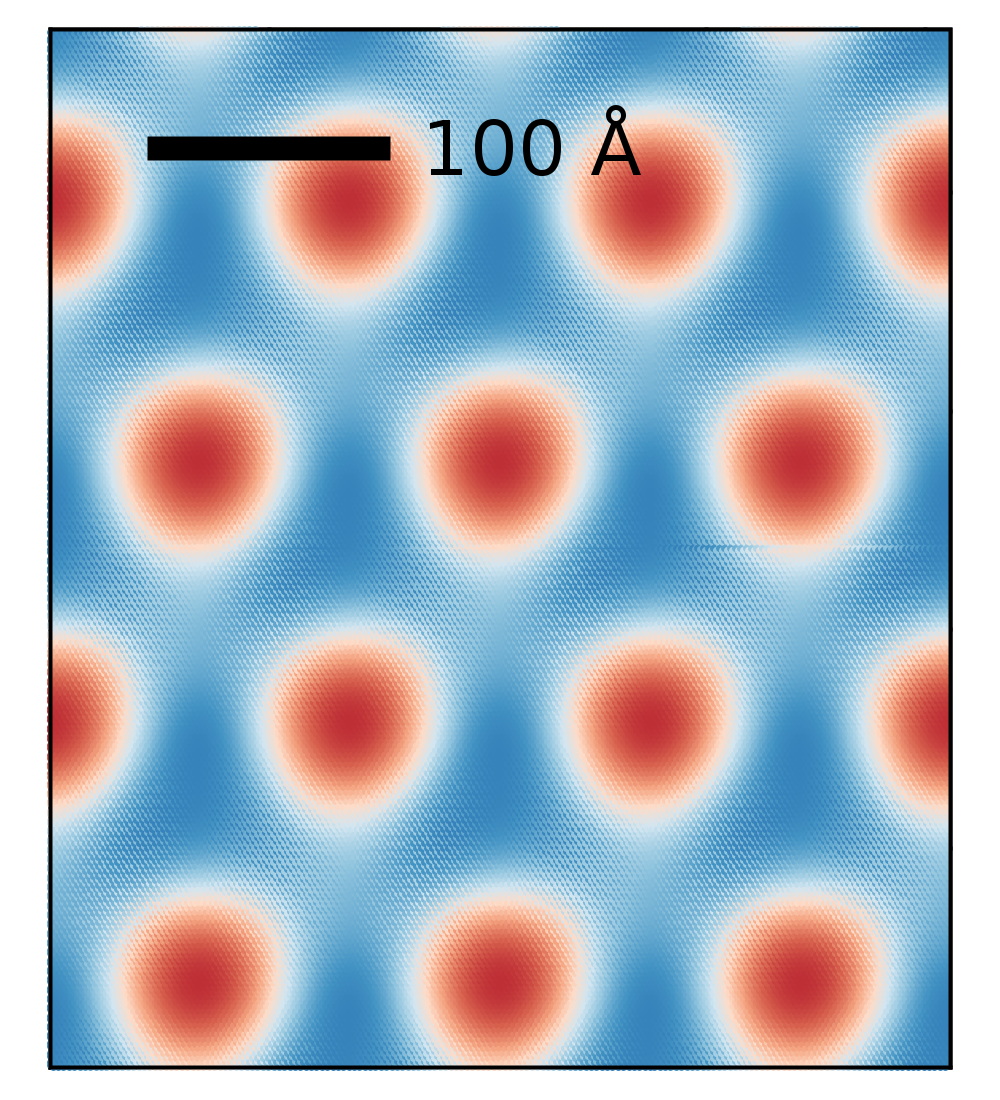}           
    \includegraphics[width=0.1533333333333333\linewidth]{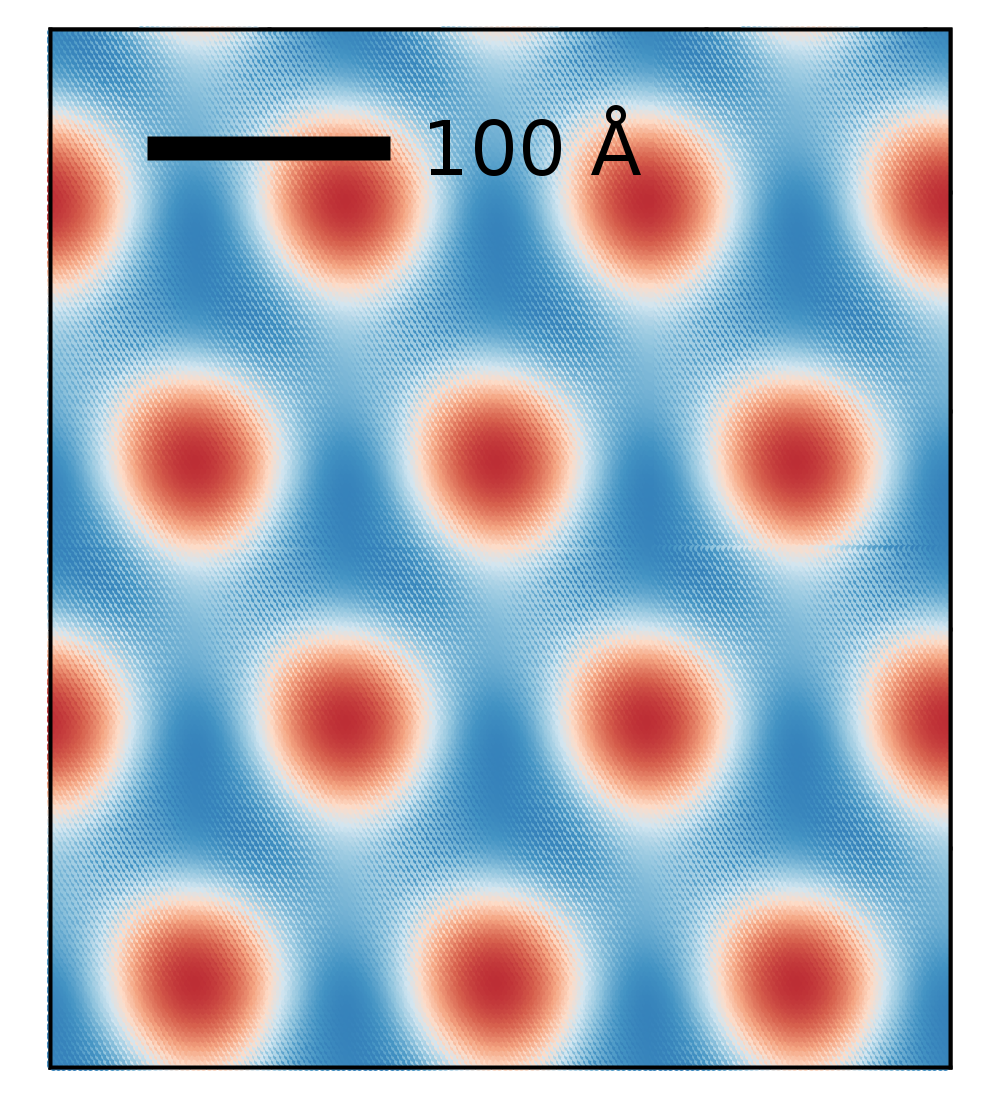}
    \includegraphics[width=0.23\linewidth]{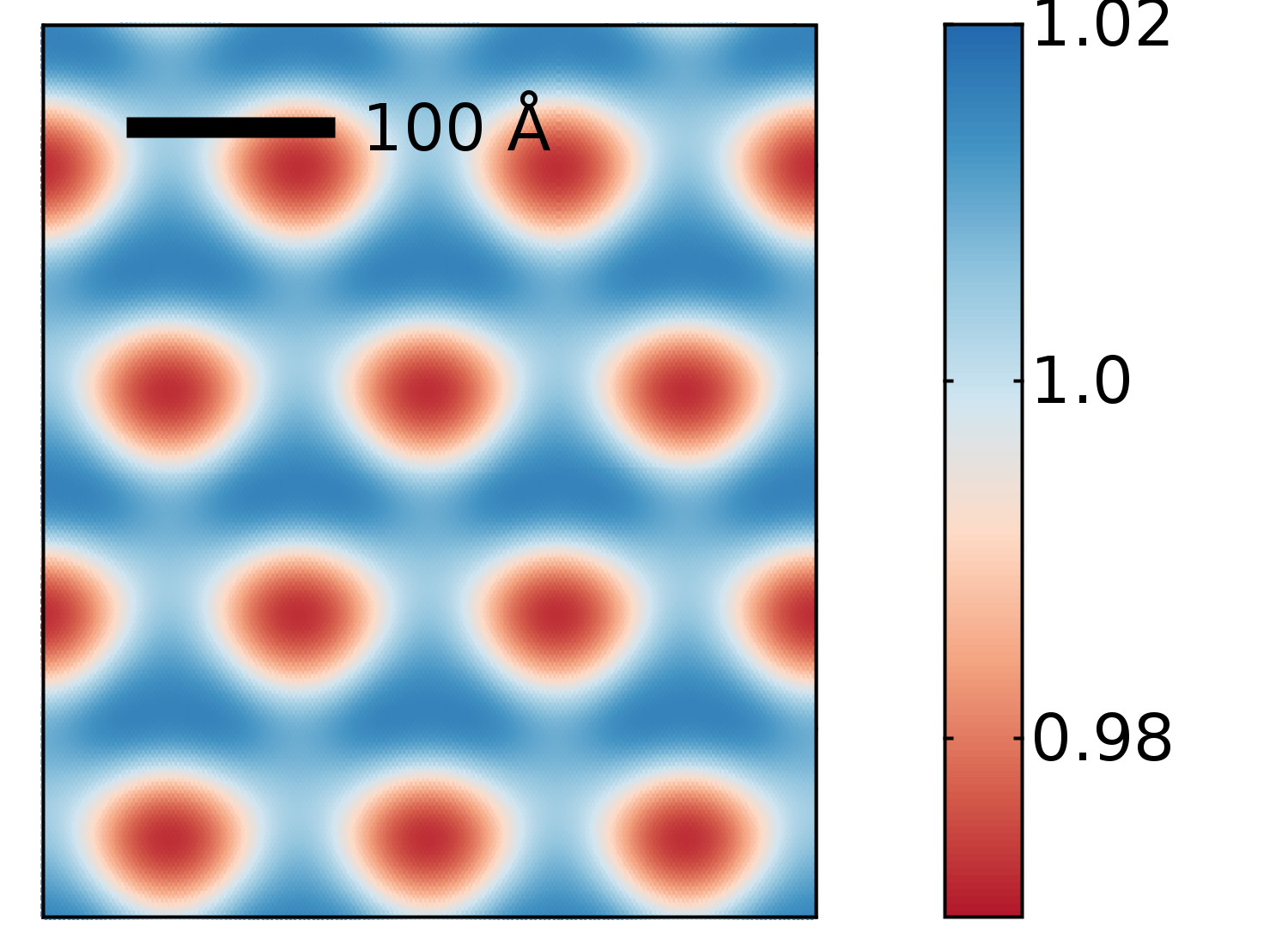}
    
    \caption{The modified TB parameters for a relaxed sample of graphene on hBN with $\theta = 0^\circ$ ($\lambda = 13.8$~nm). From left to right the on-site potential $v$, and the hopping parameters $t_1$, $t_2$ and $t_3$. The color bars are in units of $t=2.7$~eV.   }  
\label{fig:tb-para}
\end{figure*}
%%%%%%%%%%%%%%%%%%%%%

%structural relaxation
The first step is the relaxation of graphene on hBN. We follow the approach of Ref.~\cite{merel2014prl}, where it was shown that moir\'e patterns can be used as a probe of interplanar interactions for graphene on hBN. We construct supercells of rotated graphene on hBN with misorientation angles $\theta$ and corresponding moir\'e patterns with period $\lambda$~\cite{*[{See Supplemental Material (url) for details about the construction of the supercells. The relation between $\theta$ and $\lambda$ is given in }] [{.}] hermann2012}.
The graphene atoms interact through the reactive empirical bond order potential REBO~\cite{rebo}, as implemented in the molecular dynamics code LAMMPS~\cite{lammps}. The hBN substrate is kept rigid, mimicking a bulk substrate. As no empirical potential for the interactions between graphene and hBN is available, we use the registry-dependent Kolmogorov-Crespi potential~\cite{kolmogorov2005registry} developed for graphite. We neglect the correction for bending introduced to describe carbon nanotubes. 
We set the ratio of C-B/C-N interactions to $30$\% with the C-N interaction twice as strong as the original C-C interaction, as this leads to better agreement with experimental results~\cite{woods2014commensurate,merel2014prl} and ab initio calculations~\cite{sachs2011adhesion, bokdam2014band}. We minimize the total potential energy by relaxing the graphene layer by means of FIRE~\cite{bitzek2006fire}, a damped dynamics algorithm. For aligned samples ($\theta=0^\circ$), this relaxation leads to significant changes in bond length along the moir\'e pattern. The degree of deformation decreases with increasing angle~\footnote{See Supplemental Material (url) for the moir\'e patterns with different rotation angels.}.

%tight binding hamiltonian
After relaxation, we use the following graphene TB-Hamiltonian.
The main idea of our method is that the TB parameters are modified as a function of a small displacement out of equilibrium of the carbon atoms. The general TB-Hamiltonian for graphene is given by:
\begin{equation}
H = - \sum_{\left<i,j\right>} t_{ij} c_{i}^\dagger c_j + \sum_{i} v_i c_{i}^\dagger c_j  ,
\label{eq:tb-hamil}
\end{equation}
where only the nearest-neighbor hopping and on-site potential are taken into account. Including next-nearest-neighbor hoppings will result in minor changes~\cite{capacitance2013}. The change in the hopping parameter $t_{ij}$ can be written as~\cite{pereira2009tight}:
\begin{equation}
t_{ij} = t \exp(-3.37 (r_{ij}/a_0 -1)),
\end{equation}
where $r_{ij}$ is the distance between atoms $i$ and $j$, $t = 2.7$~eV is the regular hopping parameter, and $a_0 = 1.42$~\AA~is the equilibrium carbon-carbon distance for graphene.
For the on-site potential $v_i$ we calculate an effective area $S$ of each carbon atom~\footnote{We project atom $i$ to the plane formed by its three neighbors $j$. }, which will be changed due to local deformations resulting in a modulated value for $v_i$: 
\begin{equation}
v_i = g_1 \frac{\Delta S}{S_0},
\end{equation}
where $g_1=4$~eV. This value corresponds to the screened deformation potential, which gives reasonable description of transport properties~\cite{ochoa2011temperature}, and is close to density functional estimates~\cite{choi2010effects}.
Figure~\ref{fig:tb-para} shows the change of the on-site potential and of the hopping parameters for a relaxed layer of graphene on hBN with rotation angle $\theta=0^\circ$. A clear periodic modulation with period $\lambda$ is found in all parameters. 

Electronic properties are calculated using the TBPM, a method based on the propagation of the wavefunction of the time-dependent Schr\"odinger equation using Chebychev polynomials ~\cite{yuan2010modeling,hams2000fast}. The correlation function $\left< \phi(0) | e^{-iHt} | \phi(0) \right>$ is calculated at each time step. The density of states (DOS) can then be obtained by a Fourier transform of these correlation function. To increase the accuracy of the electronic calculations the supercells are repeated so that the total system consists of $\sim6000\times6000$ carbon atoms. 

%%%%%%%%%%%%figure relaxed vs unrelaxed
\begin{figure}
\subfloat[\label{fig:relaxed_vs_unrelaxed}]{%
      \includegraphics[width=0.23\textwidth]{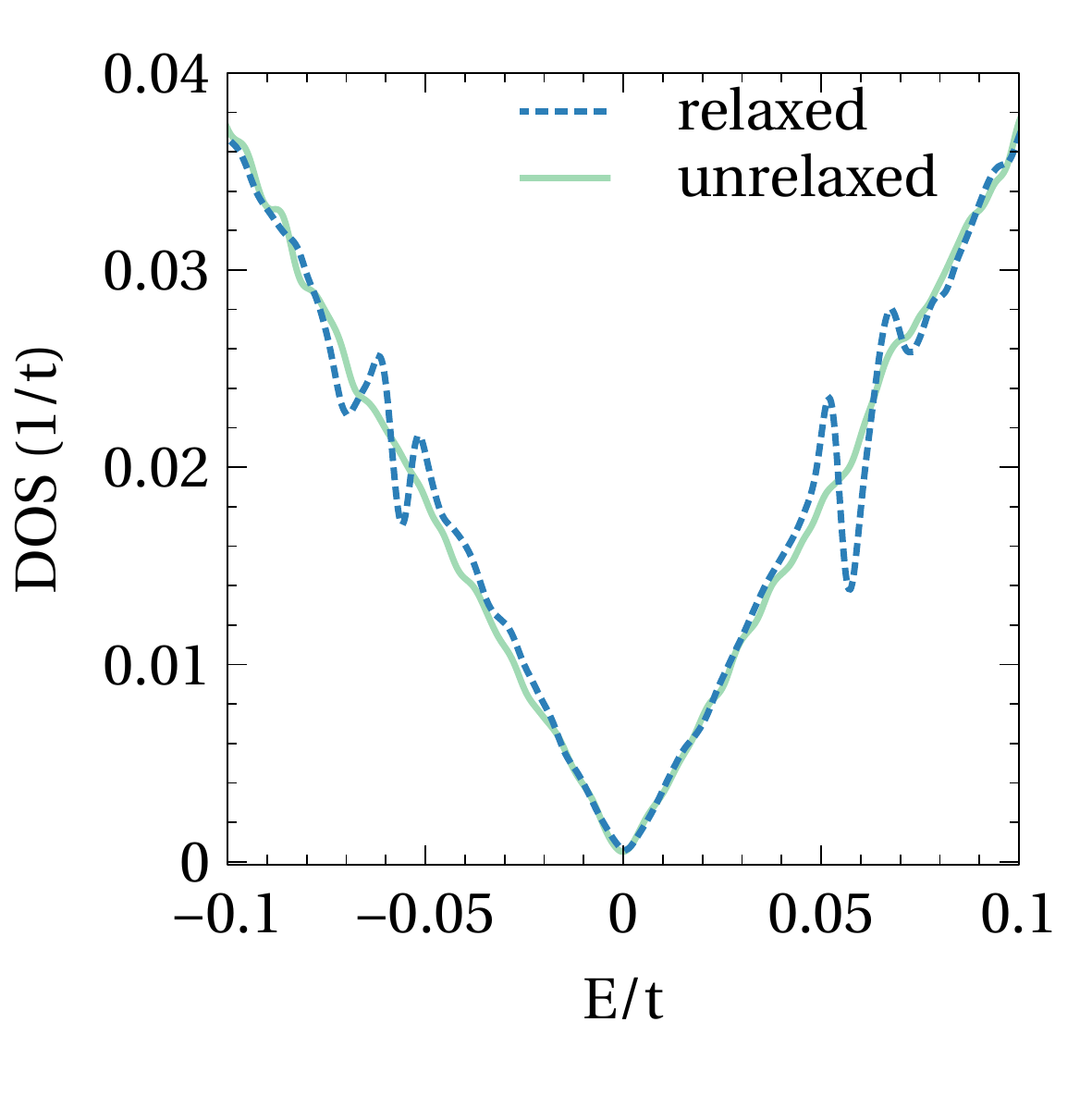}
}
\subfloat[\label{fig:rotation_dos}]{%
      \includegraphics[width=0.23\textwidth]{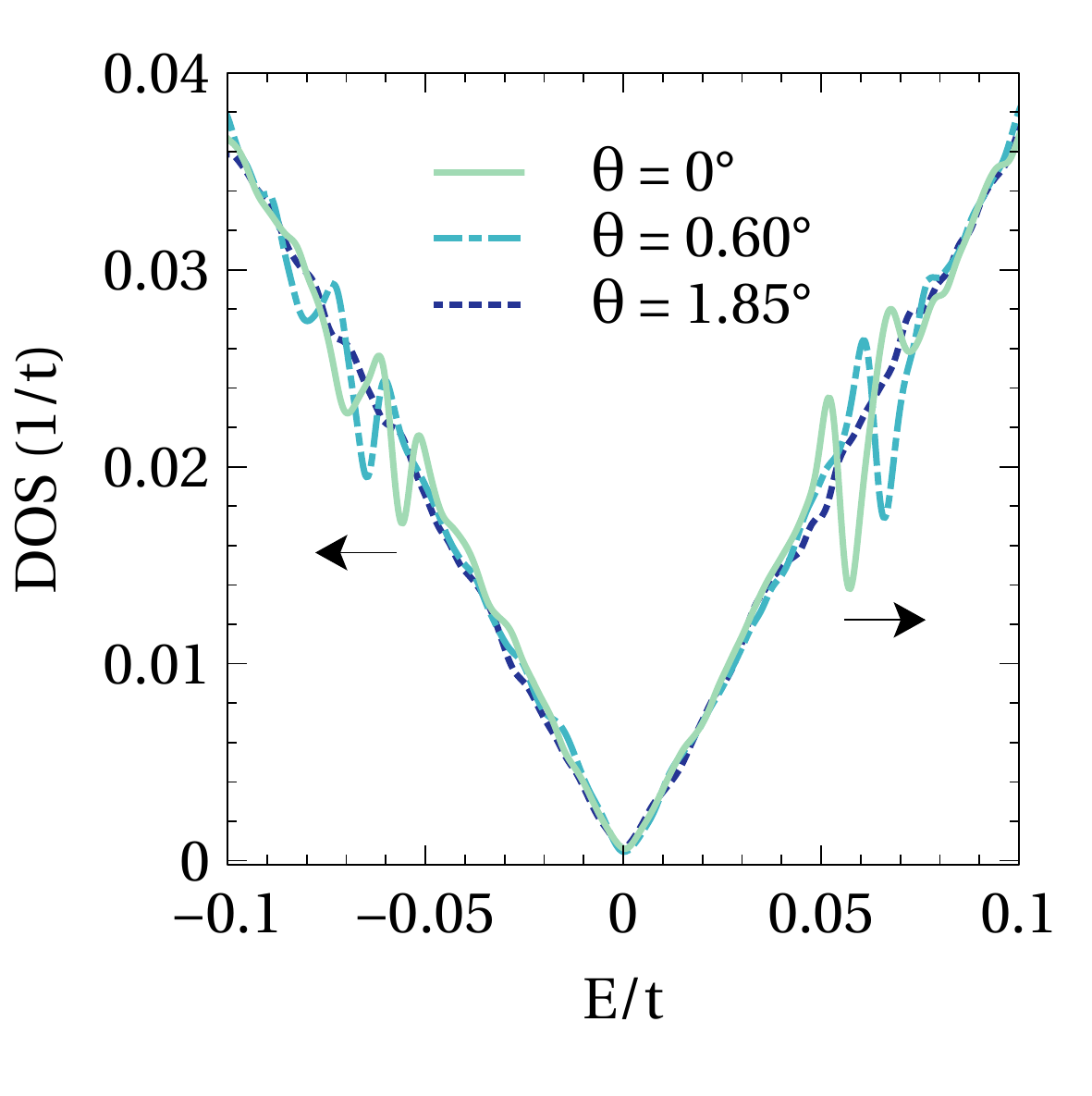}
}
    \caption{(a) Numerical results for the DOS of unrelaxed and relaxed graphene. (b) DOS for different angles $\theta$. The extra cones move outward, indicated by the arrows, and disappear for large angles. The corresponding moir\'e lengths $\lambda$ are: $13.8$~nm, $11.9$~nm and $6.7$~nm respectively. }  
\label{fig:relaxed_rotated}
\end{figure}
%%%%%%%%%%%%%
The first step to validate our method is to compare the DOS of pristine graphene (unrelaxed) to that of graphene on hBN after energy minimization (relaxed), as shown in Figure~\ref{fig:relaxed_vs_unrelaxed}. 
Secondary Dirac cones appear at both the electron and hole side, as seen in experiments~\cite{xue2011scanning, yankowitz2012emergence,hunt2013massive, ponomarenko2013cloning, dean2013hofstadter}. The positions depend on the reciprocal lattice vector $\textbf{G}$ of the moir\'e pattern, and is given by $E_D = \pm \hbar v_F \left| \textbf{G} \right| / 2=\pm 2 \pi \hbar v_F / (\sqrt{3}\lambda)$ where $v_F$ is the Fermi velocity~\cite{park2008new,park2008anisotropic,wallbank2013generic}. 
Due to the substrate the depth of the extra cones is asymmetric and highly dependent on the value of the on-site potential. The position of the extra Dirac cones will change with misorientation angle $\theta$ as $\lambda$ depends on $\theta$~\cite{hermann2012}. Figure~\ref{fig:rotation_dos} shows how small angular variations shift the extra cones. The effect of the relaxation decreases with increasing $\theta$, meaning that the differences of the DOS also become negligible for large $\theta$.

%%%%%%%%%%%%%%%quasieigenstates
\begin{figure}[ht]
    \includegraphics[width=0.20\textwidth]{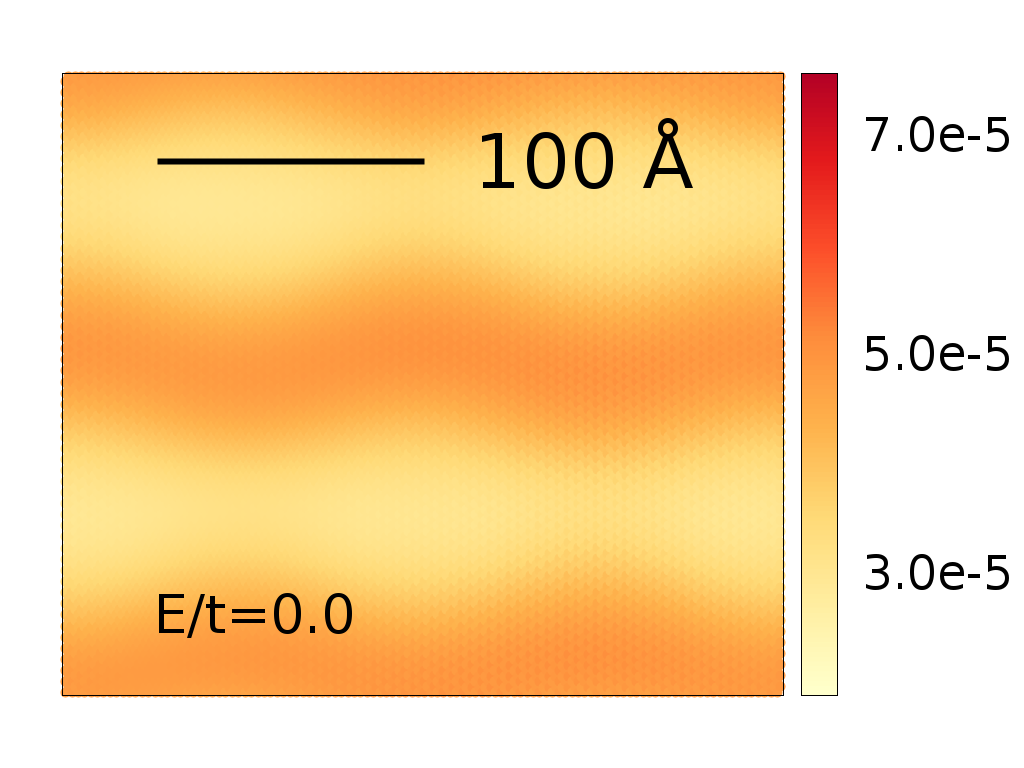}%125
    \includegraphics[width=0.20\textwidth]{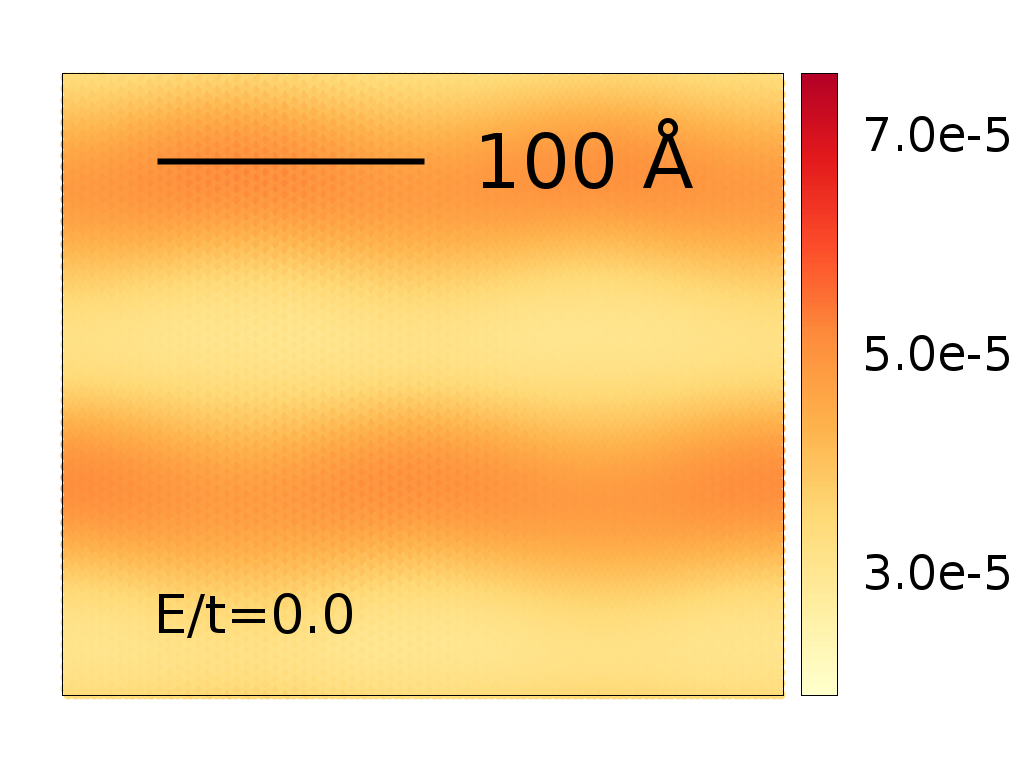}
    \includegraphics[width=0.20\textwidth]{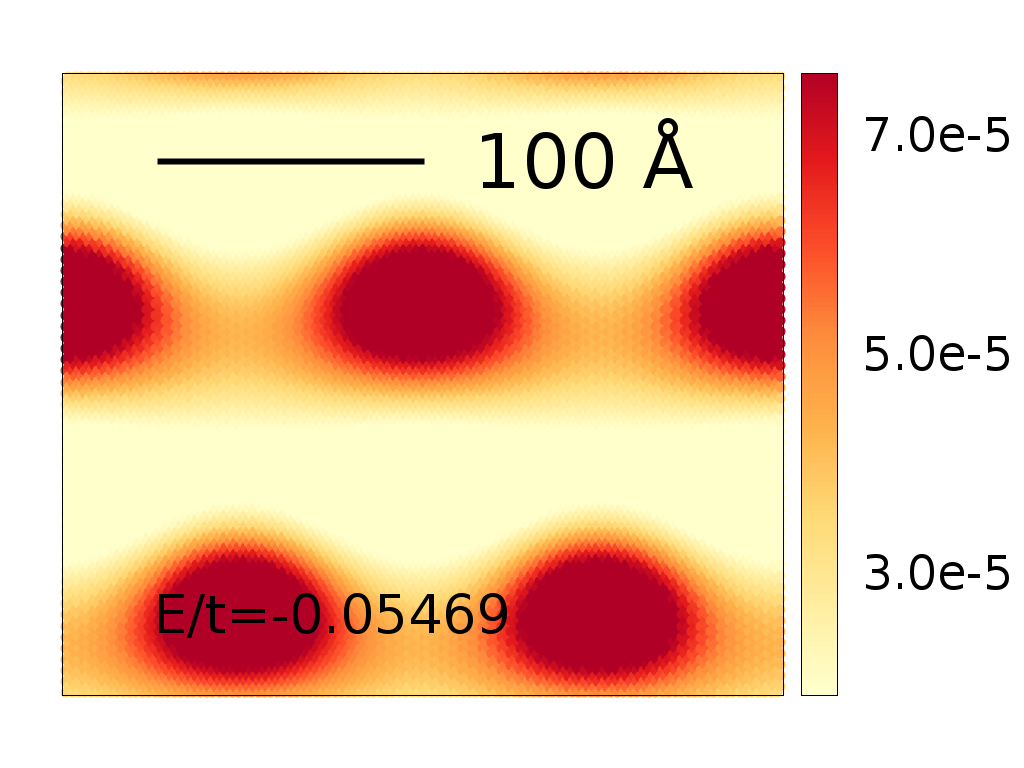}%111
    \includegraphics[width=0.20\textwidth]{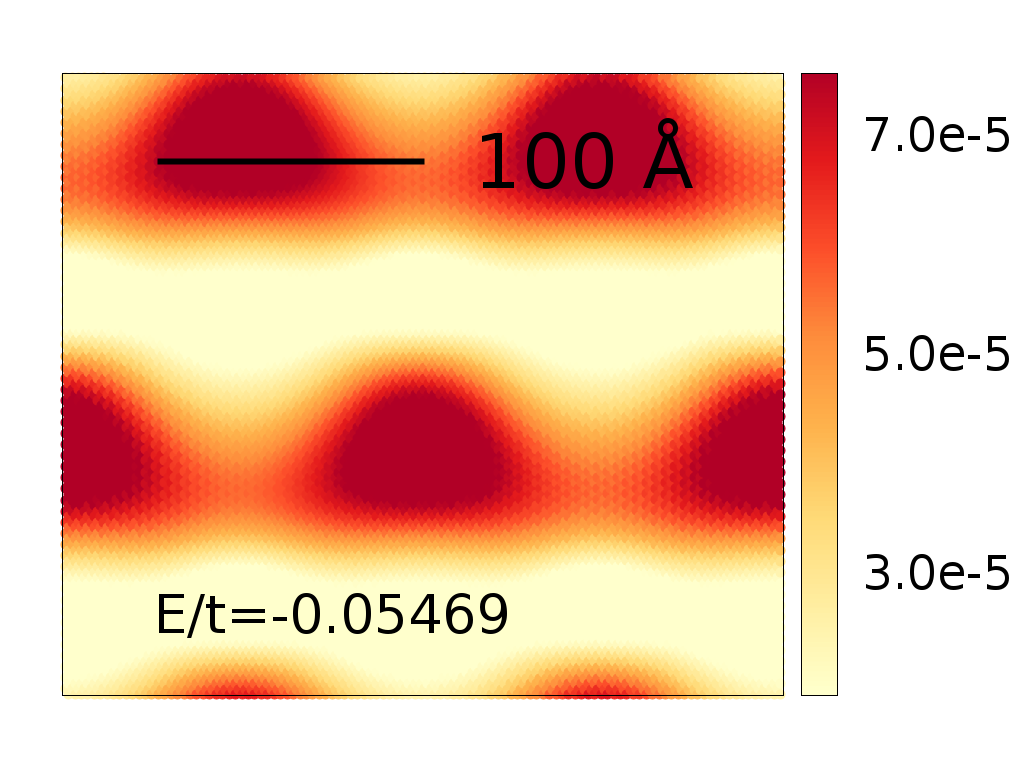}
    \includegraphics[width=0.20\textwidth]{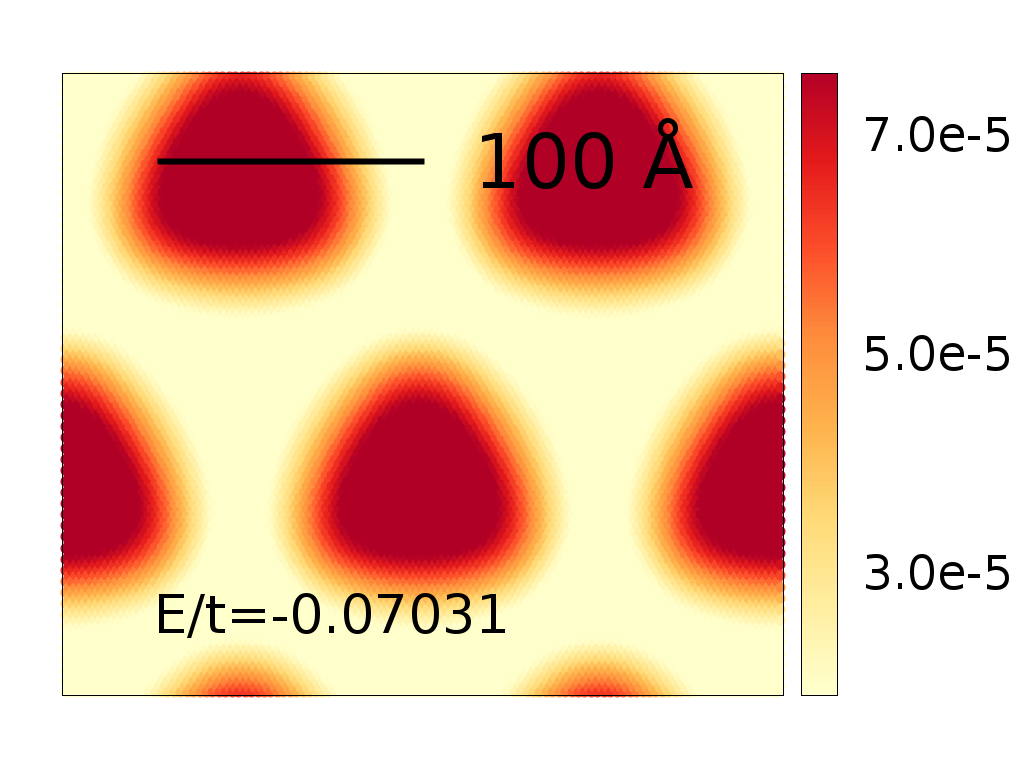}%107
    \includegraphics[width=0.20\textwidth]{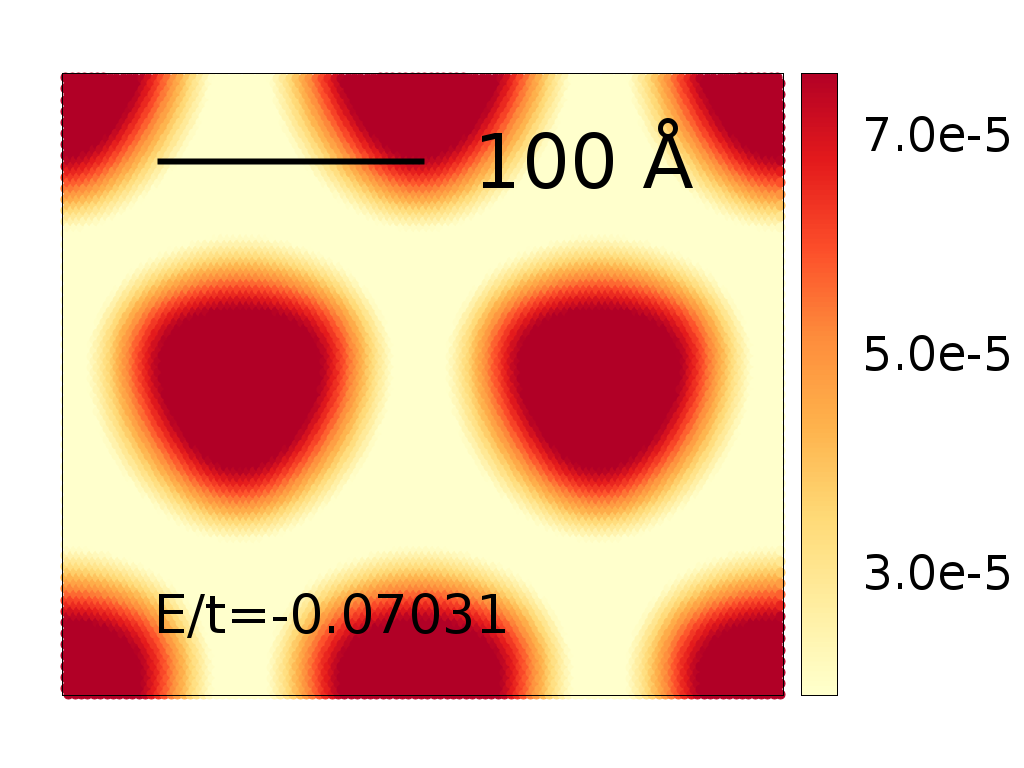}
    \caption{Amplitude of the quasi-eigenstates for different energies for $\theta=0^{\circ}$. The left-hand panels show sublattice A and the right-hand panels show sublattice B. For energies closer to the extra Dirac cones a clear moir\'e pattern can be distinguished. Only roughly one thousands of the system is shown. }
\label{fig:quasieigenstates}
\end{figure}
The real-space distribution of eigenstates can be compared with the LDOS images obtained from STM measurements. In general it is hard to obtain the eigenstates corresponding to a TB Hamiltonian of a system with millions of atoms. We obtain the so called quasi-eigenstates~\cite{yuan2010modeling}, which are close to the real eigenstates, by using the TBPM. Figure~\ref{fig:quasieigenstates} shows the amplitude of some of these quasi-eigenstates close to the additional Dirac cones. The hBN substrate breaks the sublattice symmetry, and therefore we plot the quasi-eigenstates separately. Some localization is found for the quasi-eigenstates. We see that for energies close to the Fermi energy the difference between amplitudes is negligible. For energies closer to the additional Dirac cones a clear moir\'e pattern can be distinguished.

\begin{figure}[th]
% \subfloat[\label{fig:g2_a}]{%
      \includegraphics[width=0.23\textwidth]{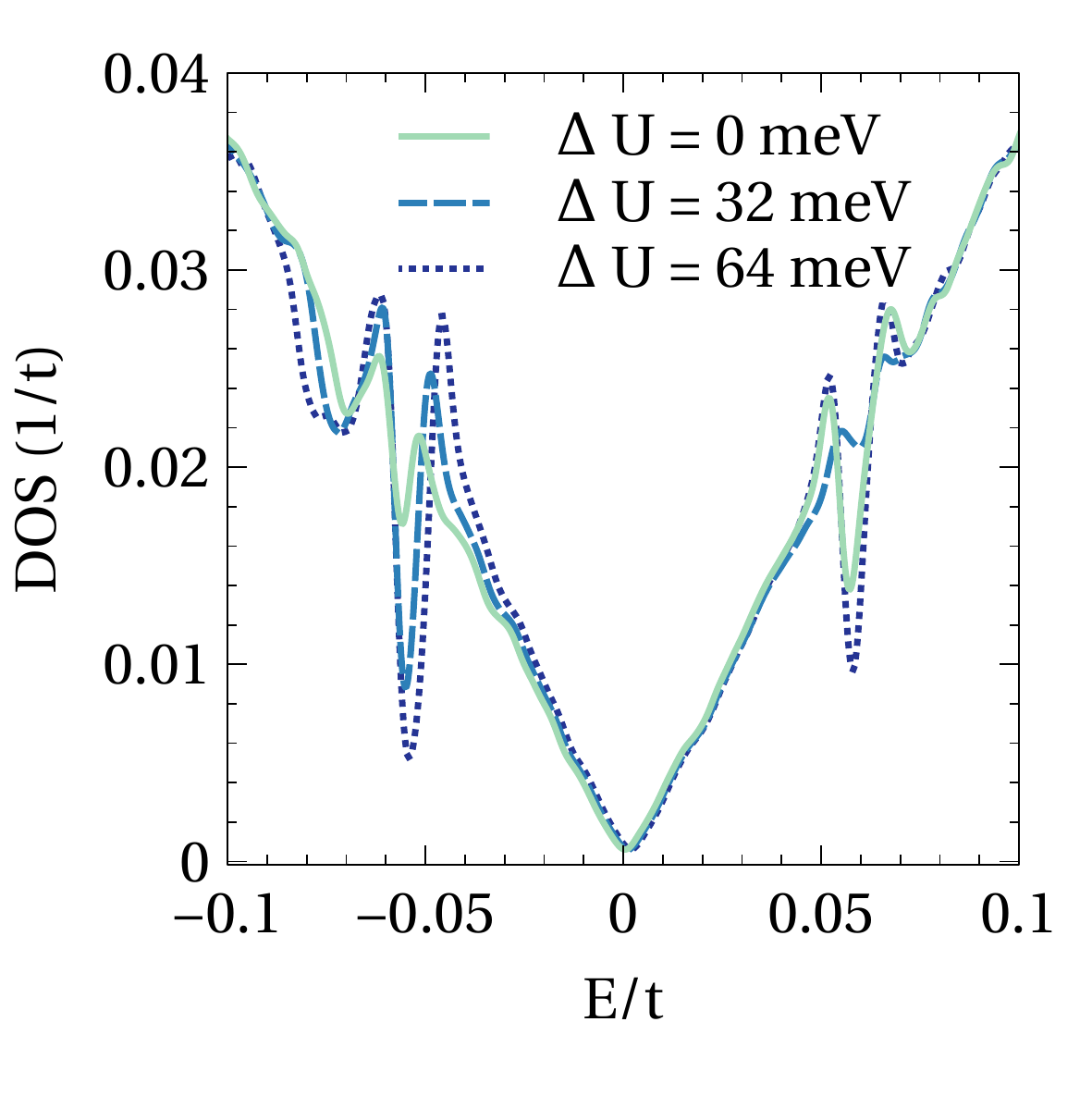}
% }
% \subfloat[\label{fig:g2_b}]{%
      \includegraphics[width=0.23\textwidth]{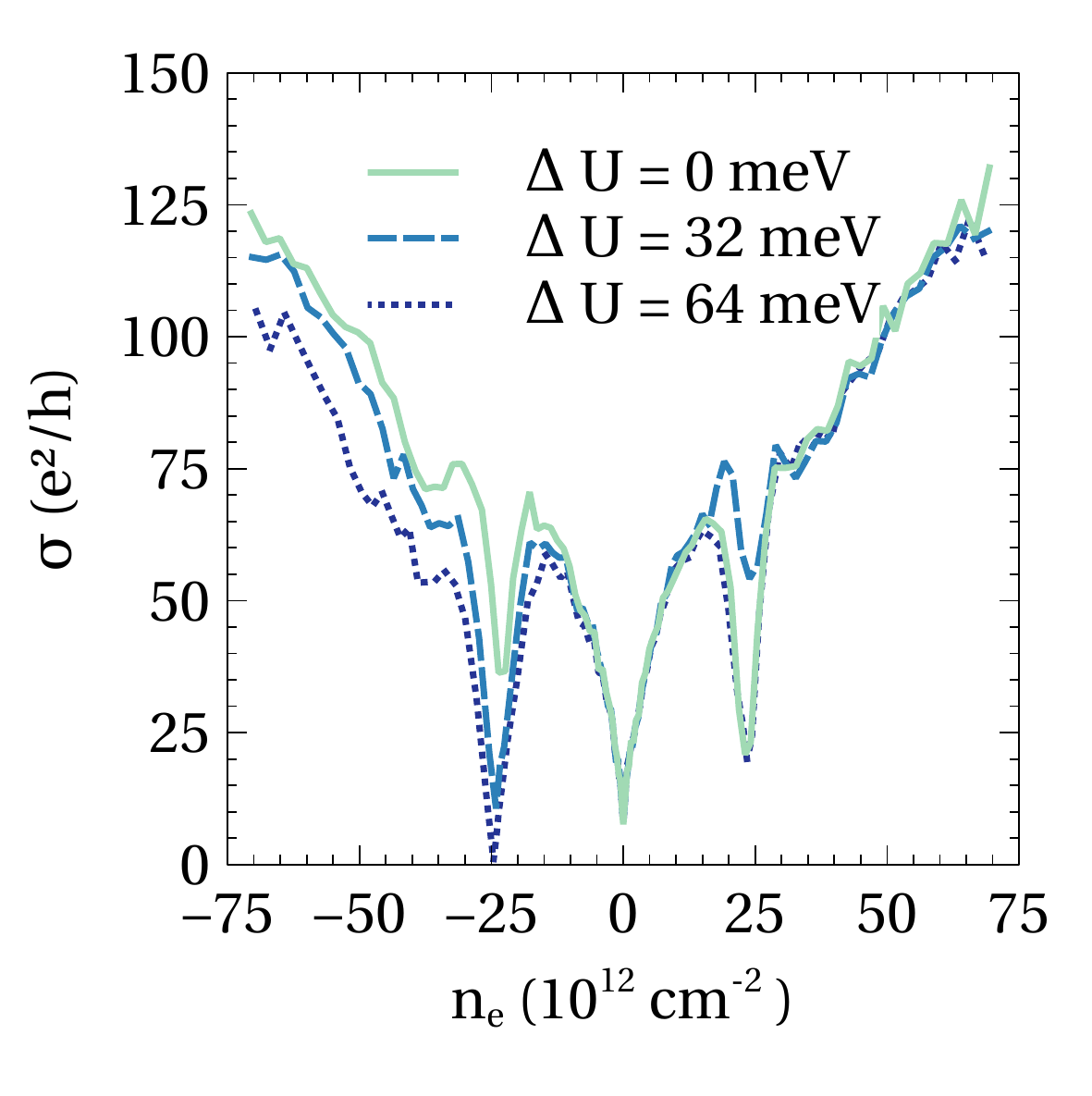}
% }
    \caption{Density of states (left) and DC conductivity $\sigma$ (right) as a function of the carrier density $n_e$ for $\theta=0^{\circ}$ and for varying $\Delta U$.  }  
\label{fig:g2}
\end{figure}

The appearance of additional Dirac cones in the DOS and the signatures of localization in the quasi-eigenstates indicate that the electronic structure is strongly influenced by relaxation. The transport measurements of DC conductivity of graphene on hBN in recent experiments~\cite{hunt2013massive, ponomarenko2013cloning, dean2013hofstadter} show clearly asymmetric drops of the conductivity at the secondary Dirac points on the hole and electron sides. The decreasing of the conductivity on the hole side is more significant, with a value even lower than the minimum conductivity at the Dirac point in Ref. ~\cite{ponomarenko2013cloning}. 
We calculate the DC conductivity by using the Kubo formalism~\cite{kubo1957} within the TBPM~\cite{yuan2010modeling}.
The results shown in figure~\ref{fig:g2} do not have such minimum on the hole side as in experiments.
It could be obtained by using interaction strength in the empirical potential used for the relaxation~\footnote{See Supplemental Material (url) for the role of the interaction strength.} much stronger than is suggested by \textit{ab initio} total energy calculations~\cite{sachs2011adhesion}. However, there is an interaction which we have not yet considered, namely the local gap opening induced by the substrate~\cite{sachs2011adhesion}.
It is known that the many-body effects can increase the gap dramatically~\cite{song2013electron}, and more accurate GW calculation gives a several times larger gap~\cite{bokdam2014band} in comparison with DFT~\cite{sachs2011adhesion}. To take into account the sublattice asymmetry due to many-body effect, we add a local gap term according to the potential difference between one site and its three neighbors as:

\begin{equation}
v'_i = v_i + \Delta v_i = v_i + \frac{g_2}{2}\left[ v_i - \frac{1}{3} \sum_{\delta = 1,2,3} v_{i+\delta} \right], 
\label{eq:g2}
\end{equation}

The strength of the local gap, which is controlled by the parameter $g_2$ in Eq.~\ref{eq:g2}, is given by the average of the potential difference between sublattices A and B $\Delta U = \left< \left| \Delta v_i \right| \right> $.  Numerical calculations of the DOS in Figure~\ref{fig:g2} show that the depth of the additional minima at energy $E_D$ can be tuned by the local gap $\Delta U$. For increasing $\Delta U$, the minimum on the hole side of the DOS becomes deeper, while the one on the electron side first disappears for small $\Delta U$ and then reappears for large $\Delta U$. Although it is very difficult to estimate $\Delta U$ accurately since there is no quantitatively accurate theory of many-body effects in graphene, we can use the one obtained by \citet{bokdam2014band}, a GW band gap of $32$ meV for incommensurable graphene on hBN with $\theta = 0^\circ$ as a reference value.  For $\Delta U = 32$ meV, we see clearly a decrease (increase) of DOS and DC conductivity ($\sigma$) at the extra Dirac point on the hole (electron) side. The transport calculation with $\Delta U = 32$~meV reproduces well the experimental observations in Ref. ~\cite{hunt2013massive, dean2013hofstadter}. On the other hand, in Ref. ~\cite{ponomarenko2013cloning}, the value of DC conductivity at the extra Dirac point on the hole side is smaller than the minimum conductivity at the Dirac point. This is only possible by using a larger $\Delta U$, for example, $\sigma$ drops to zero by doubling $\Delta U$ as $64$ meV. Our numerical results suggest that the experimentally observed insulating state at the extra Dirac point  on the hole side~\cite{hunt2013massive, ponomarenko2013cloning, dean2013hofstadter} is a signature of strong local gap induced by many-body effects.

%magnetic fields%

\begin{figure}[th]
% \subfloat[\label{fig:magnetic_field_a}]{%
      \includegraphics[width=0.35\textwidth]{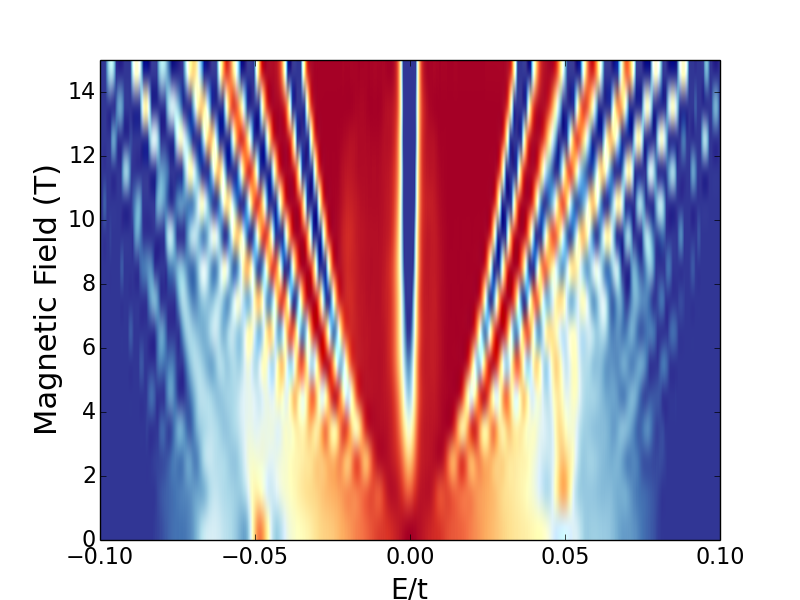}
% }\\
% \subfloat[\label{fig:magnetic_field_b}]{%
      \includegraphics[width=0.35\textwidth]{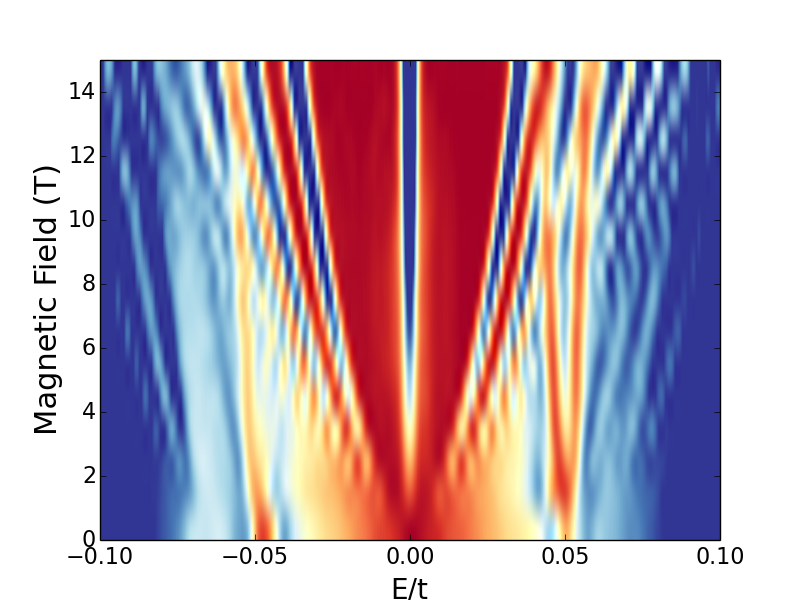}
% }
    \caption{The density of states for varying magnetic field with (top) $\Delta U = 32$~meV and (bottom) $\Delta U = 64$~meV. Strong extra peaks at both electron and hole side are observed, which for higher magnetic fields split into two. }  
\label{fig:magnetic_field}
\end{figure}

In the presence of a perpendicular magnetic field, the quantization of the energy eigenstates leads to discrete Landau levels. The modulation induced by the moir\'e patterns, splits the flat Landau bands of pristine graphene into minibands, the so called "Hofstadter butterfly spectrum" which has been conformed in several recent experiments~\cite{hunt2013massive, ponomarenko2013cloning, dean2013hofstadter}. In order to verify the splitting of the Landau levels in our TB model, we show the contour plot of DOS as a function of magnetic field strengths in Figure~\ref{fig:magnetic_field}.  For both $\Delta U = 32$ and $\Delta U = 64$~meV, there is a clear splitting of the Landau levels with increasing magnetic field, and the splitting becomes more clear when the stronger local gap term is included.

%optical conductivity

\begin{figure}[th]
\mbox{
% \subfloat[\label{fig:optical_conductivity_a}]{%
      \includegraphics[width=0.24\textwidth]{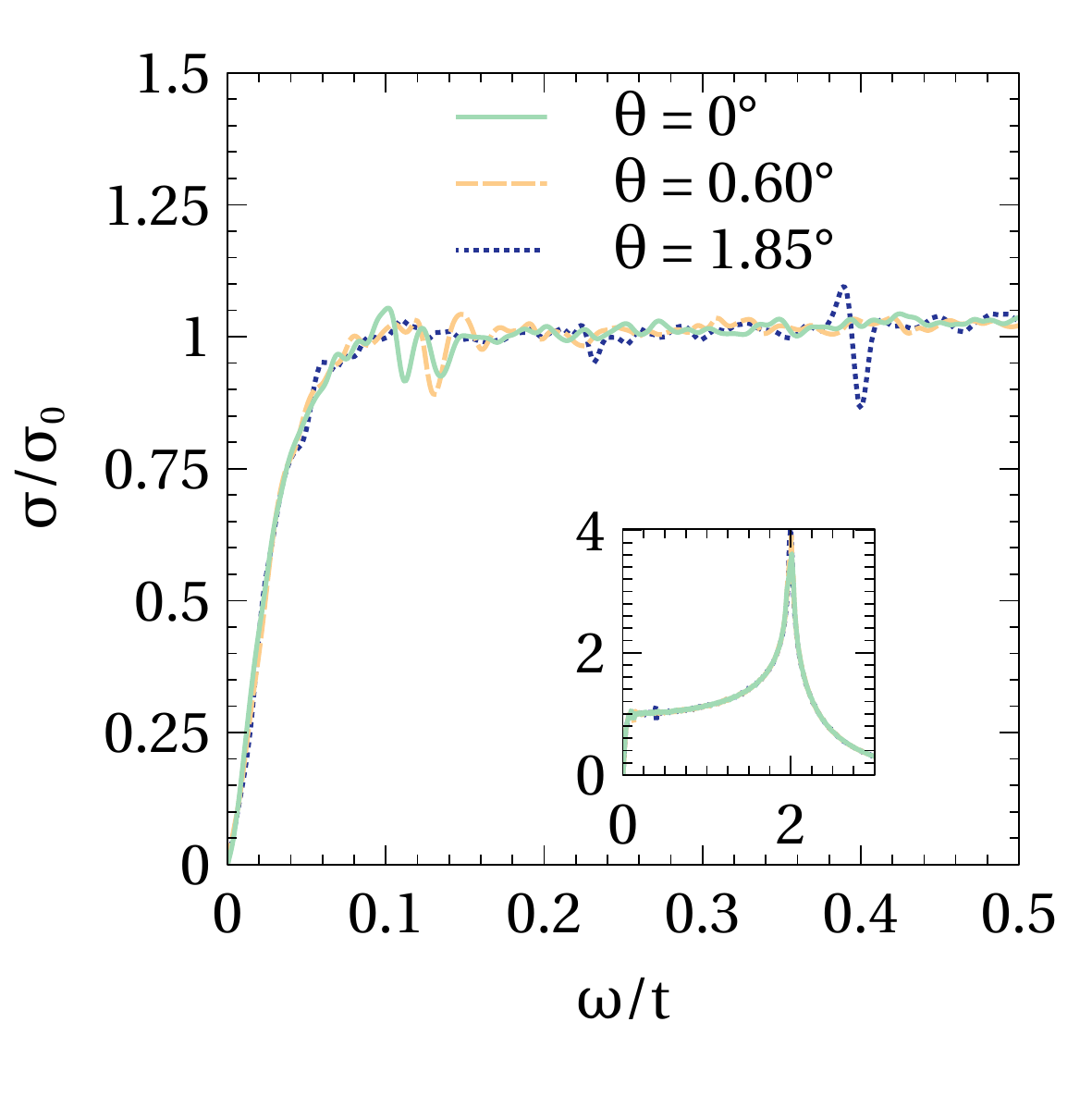}
% }
% \subfloat[\label{fig:optical_conductivity_b}]{%
      \includegraphics[width=0.24\textwidth]{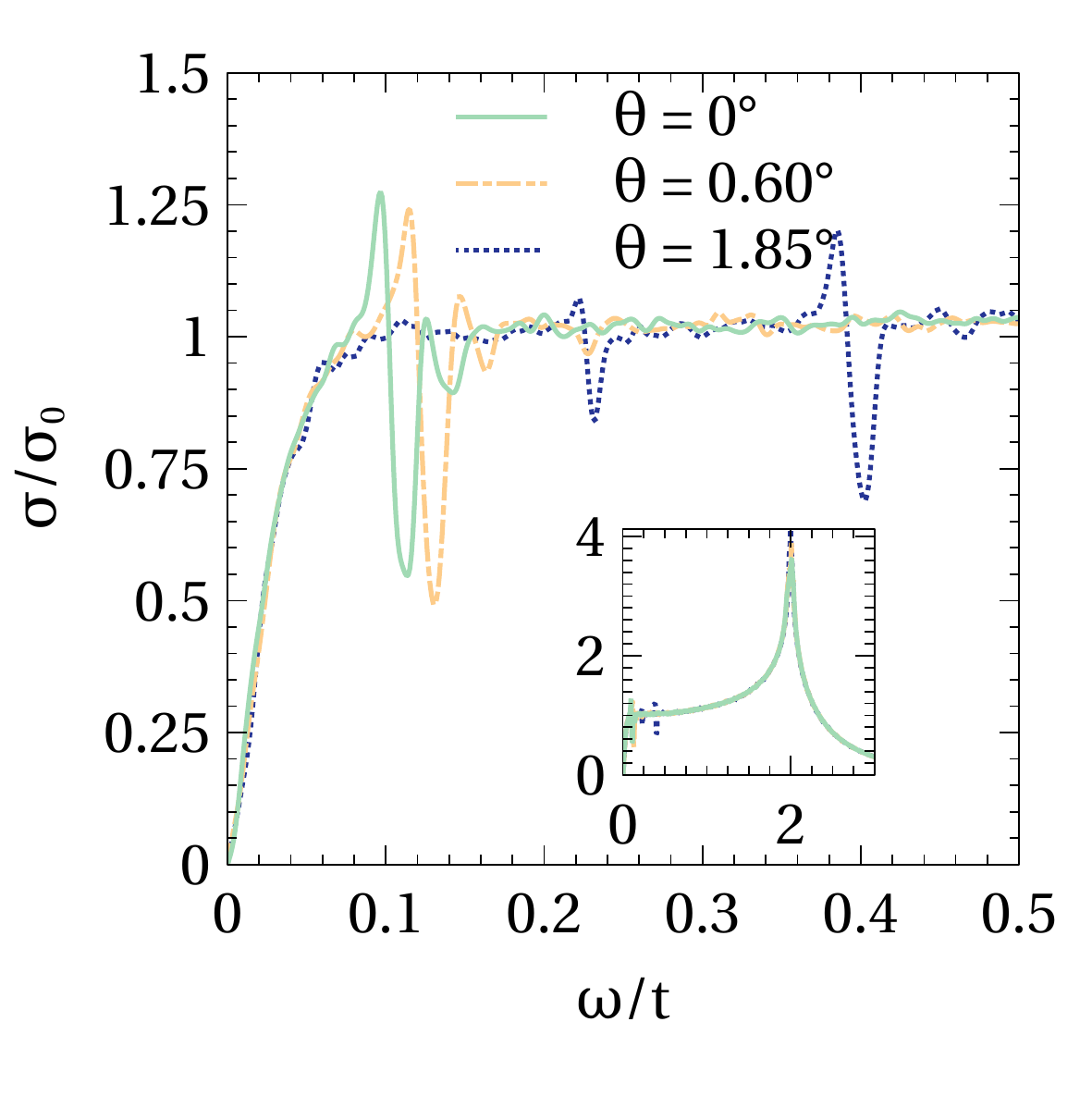}
% }
}
    \caption{Optical conductivity $\sigma$ for various angles $\theta$ with (left)  $\Delta U = 32$~meV and (right) $\Delta U = 64$~meV. $\sigma_0 = \pi e^2 / 2h$ is the universal optical conductivity of graphene. Inset: the optical conductivity for larger scales. }  
\label{fig:optical_conductivity}
\end{figure}

Another quantity of great experimental and practical interest is the optical conductivity, that we calculate by using the TBPM~\cite{yuan2010modeling,yuan2011optical}. Due to the presence of moir\'e pattern, we expect that there should be signatures of the extra Dirac cones in the optical spectrum. Figure~\ref{fig:optical_conductivity} shows the optical spectrum $\sigma$ of graphene on hBN with three different orientation angles $\theta$. 
For high energies the enhanced peak around $\omega=2t$, resulting from the optical transition between Van Hove singularities at $E=\pm t$, is similar to pristine graphene. Futhermore, there are additional peaks at photon energy about $\omega = 2 \left|E_D\right|$  (around  $0.1 \sim 0.2$~t, depending on the angle $\theta$ ), corresponding to the optical transitions between the peak states around the extra Dirac points on the hole and electron sides. The amplitudes of these peaks increase significantly with larger local gap term. It is known that the optical conductivity of graphene for visible light has a universal value, our results with moir\'e patterns indicate that the optical conductivity becomes tunable by changing the relative orientations between graphene and its hBN substrate. 

%conclusions
To conclude, we have shown that merely taking into account the periodic modulation in graphene caused by a substrate is enough to describe new features in the electronic and optical properties of graphene. The many-body enhancement of the local energy gap is crucially important to reproduce the experimentally observed insulating state at the extra Dirac point on the hole side. We also show that the optical conductivity of graphene is tunable by varying the relative orientations between graphene and its hBN substrate. The presented approach for the construction of the TB model is not limited to graphene-BN heterostructures, but can be used for graphene with other substrates, such as Ru and Cu, and can be extended to include various types of disorder.

This work is part of the research program of the Foundation for Fundamental Research on Matter (FOM), which is part of the Netherlands Organisation for Scientific Research (NWO). The research leading to these results has received funding from the European Union Seventh Framework Programme under grant agreement No. 604391 Graphene Flagship and was supported by the ERC Advanced Grant No. 338957 FEMTO/NANO.

\bibliography{bibliography}

%\appendix
%%faster compilation without appendix
%\include{appendix}

\end{document}